\documentclass[reprint,
superscriptaddress,
 amsmath,amssymb,
 aps
]{revtex4-1}

\usepackage{footmisc}
\usepackage{dcolumn}
\usepackage{bm}
\usepackage{xcolor} 
\usepackage{hyperref}       
\usepackage{url}            
\usepackage{booktabs}       
\usepackage{amsfonts}       
\usepackage{nicefrac}       
\usepackage{microtype}      
\usepackage{amsmath}
\usepackage{enumerate}   
\usepackage{mathtools}
\mathtoolsset{showonlyrefs=true}

\usepackage{graphicx} 
\usepackage{float}
\graphicspath{{Figures/}} 

\usepackage{xcolor}

\newcommand{\probp}[1]{p\left( #1 \right)}
\newcommand{\probq}[1]{q\left( #1 \right)}

\newcommand{\sampleX}{\mathbf{x}}

\newcommand{\sampleU}{\mathbf{u}}
\newcommand{\probU}[1]{p_{\mathbf{u}}\left( #1 \right)}

\newcommand{\sampleZ}{\mathbf{z}}
\newcommand{\probqphi}[1]{q_\phi\left(#1 \right)}

\DeclareMathOperator*{\argmax}{arg\,max}
\DeclareMathOperator*{\argmin}{arg\,min}

\begin{document}

\title{Likelihood-free inference of experimental
Neutrino Oscillations\\
using Neural Spline Flows}

\author{Sebastian Pina-Otey}
\email{pinas@aia.es}
\affiliation{Aplicaciones en Informática Avanzada (AIA), Sant Cugat del Vall\`{e}s (Barcelona), Spain}
\affiliation{Institut de F\'{i}sica d\'{}Altes Energies (IFAE) - Barcelona Institute of Science and Technology (BIST), Bellaterra (Barcelona), Spain}

\author{Federico S\'{a}nchez}
\affiliation{University of Geneva, Section de Physique, DPNC, Geneva, Switzerland}

\author{Vicens Gaitan}
\affiliation{Aplicaciones en Informática Avanzada (AIA), Sant Cugat del Vall\`{e}s (Barcelona), Spain}

\author{Thorsten Lux}
\affiliation{Institut de F\'{i}sica d\'{}Altes Energies (IFAE) - Barcelona Institute of Science and Technology (BIST), Bellaterra (Barcelona), Spain}

\date{\today}

\begin{abstract}
In machine learning, likelihood-free inference refers to the task of performing an analysis driven by data instead of an analytical expression. We discuss the application of Neural Spline Flows, a neural density estimation algorithm, to the likelihood-free inference problem of the measurement of neutrino oscillation parameters in Long Baseline neutrino experiments. A method adapted to physics parameter inference is developed and applied to the case of the disappearance muon neutrino analysis at the T2K experiment.  
\end{abstract}

\maketitle

\section{Introduction}

First indications of neutrino oscillations using atmospheric neutrinos were presented by the SuperKamiokande experiment in Japan in 1998\cite{Fukuda:1998mi}. The confirmation with solar neutrinos was performed in 2002 by the SNO experiment \cite{Ahmad:2001an} and the check of the atmospheric oscillation with a human-made beam experiments at K2K \cite{Ahn:2006zza} for atmospheric oscillations in 2006 and at Kamland \cite{Araki:2004mb} for solar neutrino oscillations in 2004. After eight very productive years, the oscillation phenomena was experimentally well established. Neutrino oscillation phenomena are the first, and so far the only, evidence of neutrinos having mass. Following the initial experiments, a successful set of measurements confirmed the oscillation picture and improved the understanding of the Pontecorvo-Maki-Nakagawa-Sakata (PMNS) neutrino flavour mixing matrix parameters \cite{Maki:1962mu}. By 2010, the measurement of a non-zero $\theta_{13}$ angle by T2K \cite{Abe:2011sj} , followed by the precision measurement from Daya-Bay
\cite{An:2012eh}, opened the potential existence of leptonic Charge-Parity (CP) symmetry violation (i.e. particles behaving differently from the antiparticles). As of today, all mixing angles have been measured \cite{Tanabashi:2018oca}, including the two mass differences between the three mass neutrino eigenstates. The remaining mixing matrix parameter to be measured are the imaginary phase responsible of the CP violation and 
the sign of one of the two neutrino mass splitting,
which determines the so called hierarchy. Both measurements are at reach for current and near future oscillation experiments. 

Statistical methods used in the oscillation analysis so far comprises both frequentist and Bayesian approaches \cite{Abe:2017vif,PhysRevLett.123.151803}. There are, however, some limitations to these methods. Both approaches are based as of today on a binned likelihood algorithm which might limit the sensitivity of the experiment and some of them impose a Gaussian dependency in some of the nuisance parameters affecting the precision of the results and
correctness of the evaluated uncertainties.
Additionally, they require very intensive CPU processing time, which is a limiting factor that reduces the flexibility of the statistical analysis and checks, and introduces strong constrains on the delivery of the results. These limitations are derived from the intrinsic difficulties of the statistical data analysis that are depicted in Sec.~\ref{Sec:Problemdefinition}, using the T2K experiment as a reference example.

In this paper we propose an alternative statistical method to overcome some of the limitations of the current methods in use. The proposed procedure is based on an un-binned likelihood inference using neural density estimators. This method has the potential of being accurate, fast and to reduce the possible bias due to the intrinsic binning in other approaches. The Neural Spline Flows, the implementation of neural density estimators we chose, has also some advantages since the Gaussian generator intrinsic to the method will facilitate the introduction of experimental errors in the distributions. We will discuss in this paper the basic concepts of the method and show the potential with a simplified example.  

\section{Problem definition and physical simulator}
\label{Sec:Problemdefinition}

Neutrino oscillation experiments search for the modification of the flavour content of a neutrino beam travelling in vacuum or matter for a certain distance. Beams are normally characterized at a near site, where the neutrino energy spectrum and flavour composition is not yet altered by oscillations. The same beam is sampled after a certain flight distance $L$. The change on the flavour composition can be determined in two different ways:
\begin{enumerate}[(i)]
\item The neutrino flavour disappearance ($P(\nu_{\alpha}\rightarrow \nu_{\alpha})$) experiments search for the disappearance of a certain neutrino flavour as function of the neutrino energy. The disappearance produces both a reduction in the flux of neutrinos of a given flavour and the distortion of the neutrino energy spectra that is observed in the distribution of the measured quantities. For example, T2K uses the muon momentum ($p_{\mu}$) and the angle with respect to the neutrino direction ($\theta_{\mu}$). 
\item The neutrino flavour appearance ($P(\nu_{\alpha}\rightarrow \nu_{\beta}), \alpha \ne \beta$), experiments search for the appearance of a neutrino flavour that is normally suppressed in the original neutrino flux. In T2K, this new flavour is the electron neutrino. The dependency of the oscillation with the neutrino energy is inferred from the momentum and the angle with respect to the neutrino direction of the electron ejected in the interaction of neutrinos with matter.   
\end{enumerate}
The neutrino flavour is determined by the flavour of the charged lepton (muon, electron or tauon) produced in charged current interactions of neutrinos with the nuclei. For the current analysis, we will concentrate on the disappearance phenomenon (i). 

In a synthetic way, the experimental number of observed neutrinos with observed properties ($\vec \theta^\text{reco}_{\nu}$) can be described by:
 \begin{align}
 \label{Eq:Nnear}
 N^\text{near}_\text{evts} ( \vec{\theta}^\text{reco}_{\nu} ) = & \int \sigma(E_{\nu}) \phi^\text{near}(E_{\nu}) P_\text{near}(\vec{\theta}^\text{reco}_{\nu}|E_{\nu}) d E_{\nu} \\
 & + \text{Back}_\text{near}(\vec{\theta}^\text{reco})
 \end{align}
for the near detector and
\begin{align}
\label{Eq:Nfar}
&N^\text{far}_\text{evts} ( \vec{\theta}^\text{reco}_{\nu} )  \\
& \phantom{N^\text{far}_\text{evts}}= \int \sigma(E_{\nu}) \phi^\text{far}(E_{\nu}) P_\text{far}(\vec{\theta}^\text{reco}_{\nu}|E_{\nu}) P_\text{osc}(E_{\nu}) d E_{\nu} \\ 
& \phantom{N^\text{far}_\text{evts}=}+ \text{Back}_\text{far}(\vec{\theta}^\text{reco})
\end{align}
for the far detector. The number of observed neutrinos depends on the cross-section ($\sigma(E_{\nu})$), the neutrino flux ($\phi^\text{far,near}(E_{\nu})$), the probability of observing the experimentally accessible quantities ($\vec \theta^\text{reco}_{\nu}$) given a neutrino energy ($ P_\text{far,near}(\vec{\theta}^\text{reco}_{\nu}|E_{\nu})$), the oscillation probability ($P_\text{osc}(E_{\nu})$) and the backgrounds observed in the detectors ($\text{Back}_\text{near,far}(\vec{\theta}^\text{reco})$).

 The experimental challenge comes from inferring the neutrino energy, $E_{\nu}$, given the experimental observable ($\vec \theta^\text{reco}_{\nu}$). The term $P_\text{far,near}(\vec{\theta}^\text{reco}_{\nu}|E_{\nu})$ encapsulates not only the detector resolution, but also the neutrino-nucleus cross-section model predictions and uncertainties. Other difficulties raise from the limited knowledge ($\approx$9\% in the latest T2K results\cite{Abe:2019vii}) of the neutrino flux ($\phi^\text{far,near}(E_{\nu})$) and of neutrino cross-sections as function of the energy ($\sigma(E_{\nu})$). The background terms  ($\text{Back}_\text{far,near}(\vec{\theta}^\text{reco})$) are normally relevant ($\approx$20\% in T2K)\cite{Abe:2016tmq} and they subsequently depend on the neutrino-nucleus cross-sections in a non trivial manner.

 Both the frequentist and the Bayesian statistical approaches \cite{Abe:2017vif,PhysRevLett.123.151803} utilize the near detector data to predict the probability density function at the far detector in absence of oscillations:
\begin{align}
\label{Eq:PDF} 
f( \vec{\theta}^\text{reco}_{\nu} | E_{\nu}) = \sigma(E_{\nu}) \phi^\text{far}(E_{\nu}) P_\text{far}(\vec{\theta}^\text{reco}_{\nu}|E_{\nu}).
\end{align}
Once determined, the conditional probability density function  $f( \vec{\theta}^\text{reco}_{\nu} | E_{\nu})$ can be used to determine the oscillation parameters ($P_\text{osc}(E_{\nu})$) by comparing it to the far detector events. Most of the experimental effort is actually devoted to the determination of this conditional probability which depends also on a large number of hidden and correlated parameters describing uncertainties in detector performances, cross-section models and the neutrino flux. Hidden parameters are marginalized or profiled in the analysis, providing the experimental result for oscillation parameters as (the posterior in the case of Bayesian approaches) probability maps.  

In the particular case of the T2K experiment, the experimentally accessible observables ($\vec{\theta}^\text{reco}_{\nu}$) are the momentum and direction of the $\mu$ lepton $(p_{\mu}^\text{reco},\theta{\mu}^\text{reco})$. Near and far detectors are able to provide also the kinematic of pions (charged and neutral) and protons, or the total released energy in the interaction, but we will ignore these capabilities to simplify the discussion. The $\mu$ lepton is produced at the interaction of the neutrino with the target nucleus. In this case the probability density function ($f( \vec{\theta}^\text{reco}_{\nu} | E_{\nu}) $) can be simplified to $p(p_\mu^\text{reco},\theta_\mu^\text{reco}|E_\nu)$. The near detector of the experiment measures the neutrino flux and tunes the model of
neutrino-nucleus interaction providing the estimation of the probability density function $p(p_\mu^\text{reco},\theta_\mu^\text{reco},E_\nu)=p(p_\mu^\text{reco},\theta_\mu^\text{reco}|E_\nu)p(E_\nu)$ together with the expected number of interactions in the far detector in absence of oscillations. The near detector provides also a dependency with free parameters in the model and a full error covariance matrix relating all of them. To simplify the exercise, we ignore the error covariance matrix in this study and assume that the probability $p(p_\mu^\text{reco},\theta_\mu^\text{reco},E_\nu)$ is implicitly known to the experiment through simulations.

The neutrino oscillation disappearance probability can be approached by the simplified two-flavour \cite{Tanabashi:2018oca} oscillation. The disappearance probability as a function of the initial energy of the neutrino $E_\nu$ is
\begin{equation}\label{Eq:POscillation}
\begin{aligned}
&p_\text{osc}(E_\nu,\theta_\text{mix},\Delta m^2)=\\
&\hspace{10ex} \sin^2\left(2 \theta_\text{mix}\right) \sin^2\left(1.27 \frac{\Delta m^2 295}{E_\nu}\right),
\end{aligned}
\end{equation}
where $295$ is the distance in kilometers between the near and far sites in the T2K experiment and $1.27$ is a scaling parameter to adjust the oscillation phase to distance in kilometers. $E_\nu$ is the neutrino energy in GeV, $\theta_\text{mix}$ the mixing angle of the two flavours and $\Delta m^2$ the difference in mass of the two mass eigenstates in eV$^2$. $\theta_\text{mix}$ and $\Delta m^2$ are the parameters governing the oscillations. 

Although the problem was simplified in the following analysis to make a proof of concept, in Sec.~\ref{Sec:RealCase} we outline how the methodology could be applied to the full complex analysis.

\subsection{ Physical Simulator } 
\label{Sec:PhysicalSimulator}

 We have simplified the problem to demonstrate the viability of the proposed method to determine the oscillation parameters using Neural Spline Flows. Event samples are generated using  the NEUT \cite{Hayato:2009zz} Monte Carlo event generator model that describes the interactions of neutrinos with nuclei. We also use a realistic neutrino flux energy spectrum provided by the T2K collaboration \cite{PhysRevD.87.012001}. With both inputs, we generate charge current quasi-elastic events (CCQE). CCQE is the most probable reaction at T2K energies, and the one dominating the statistical sensitivity of the experiment, where the neutrino transforms into a muon exchanging a neutron into a proton ($\nu+n\to \mu + p$). To simplify, we ignore other reaction channels and potential backgrounds, and also detector effects. The generator provides n-tuples of events weighted according to their probability as function of neutrino energy and angle and momentum of the muon. 

\section{Methodology}

Evaluating the density of high-dimensional data has become an important task for unsupervised machine learning in recent years. Neural density estimation proposes a solution using neural networks by learning an estimation $\probqphi{\sampleX}$ of the exact \textit{target density} $\probp{\sampleX}$ from samples $\sampleX\sim \probp{\sampleX}$. In this work, this task is performed through Neural Spline Flows, a specific implementation of the more general Normalizing Flows methods, presented in Sec.~\ref{Sec:NSFTheory}. We explain how the density estimation at the near detector is combined with the analytical formula for neutrino oscillation for the far detector to obtain the likelihood used to perform Bayesian inference for the T2K experiment in Sec.~\ref{Sec:NSFatT2K}. Additionally, an alternative way of computing the posterior is explained in Sec.~\ref{Sec:unbinnedlklh}, based on the standard histogram approach but tweaked in order to attain the limit of un-binned likelihood, closely related to what is obtained by Neural Spline Flows.

\subsection{Neural density estimation using Neural Spline Flows}\label{Sec:NSFTheory}

Consider a simulator which can produce samples $\sampleX\sim \probp{\sampleX}$ over the real D-dimensional space $\mathbb{R}^D$ from the target density. Consider also $\probqphi{\sampleX} = \probq{\sampleX;\phi}$ a flexible family of density functions parametrized by $\phi$, i.e., $\probqphi{\sampleX} \geq 0$ for all $\sampleX,\phi$, and $\int \probqphi{\sampleX} \; d\sampleX = 1$ for all $\phi$. Then, if we want to approximate $\probp{\sampleX}$ through $\probqphi{\sampleX}$, we optimize the parameters $\phi$ by maximizing the log-likelihood of the approximated density under simulated data from the target distribution, 
\begin{align}\label{Eq:objective}
    L = \frac{1}{N} \sum_{i=1}^N \log(\probqphi{\sampleX_i})  \text{ with } \sampleX_i\sim \probp{\sampleX}.
\end{align}

This objective function for the optimization procedure is equivalent to minimizing the Kullback-Leibler divergence between the target and approximated density,
\begin{align}\label{Eq:KLdivergence}
    D_{\text{KL}}\big(\probp{\sampleX}\| \probqphi{\sampleX}\big) = \int \probp{\sampleX} \log \left( \frac{\probp{\sampleX}}{\probqphi{\sampleX}}\right) \;d\sampleX,
\end{align}
which is always non-negative and only equal to 0 if both densities are equivalent. Let us proof quickly this statement:
\begin{align}
    &\argmin_\phi D_{\text{KL}}\big(\probp{\sampleX}\| \probqphi{\sampleX}\big)\\ 
    & \phantom{\argmin_\phi D_{\text{KL}}} = \argmin_\phi \int \probp{\sampleX} \log \left( \frac{\probp{\sampleX}}{\probqphi{\sampleX}}\right) d\sampleX\\
    & \phantom{\argmin_\phi D_{\text{KL}}}= \argmin_\phi -\int \probp{\sampleX} \log \probqphi{\sampleX} \;d\sampleX\\
    & \phantom{\argmin_\phi D_{\text{KL}}}= \argmax_\phi \int \probp{\sampleX} \log \probqphi{\sampleX} \;d\sampleX\\
    & \phantom{\argmin_\phi D_{\text{KL}}}\approx \argmax_\phi \sum \log \probqphi{\sampleX} \text{ with } \sampleX\sim \probp{\sampleX}, 
\end{align}
where in the last step we have approximated the expected value of $\log \probqphi{\sampleX}$ with respect to $\probp{\sampleX}$ by the finite expected value. Hence, maximizing the objective function Eq.~\eqref{Eq:objective} is equivalent to minimizing the KL-divergence Eq.~\eqref{Eq:KLdivergence} between the distributions, i.e., approximating $p$ by $q_\phi$.

\textit{Normalizing flows} are a mechanism of constructing such flexible probability density families $\probqphi{\sampleX}$ for continuous random variables. A comprehensive review on the topic can be found in \cite{Papamakarios2019NormalizingFF}, from which a brief summary will be shown in this Section on how normalizing flows are defined, and how the parameters $\phi$ are obtained, together with a specific implementation, the \textit{Neural Spline Flows} (NSF) \cite{Durkan2019NeuralSF}.

Consider a random variable $\sampleU$ defined over $\mathbb{R}^D$, with known probability density $\probU{\sampleU}$. A normalizing flow characterizes itself by a \textit{transformation} $T$ from another density $\probp{\sampleX}$ of a random variable $\sampleX$ defined also over $\mathbb{R}^D$, the target density, to this known density, via
\begin{align}
    \sampleU = T(\sampleX)\text{, with }\sampleX\sim\probp{\sampleX}.
\end{align}
The density $\probU{\sampleU}$ is known as \textit{base density}, and has to satisfy that it is easy to evaluate (e.g., a multivariate $D$-dimensional normal, as will be chosen through this work, or a uniform distribution in dimension $D$). The transformation $T$ has to be invertible, and both $T$ and $T^{-1}$ have to be differentiable, i.e., $T$ defines a diffeomorphism over $\mathbb{R}^D$.

This allows us to evaluate the target density by evaluating the base density using the change of variables for density functions,
\begin{align}
    \probp{\sampleX} = \probU{T(\sampleX)} |\det J_T(\sampleX)|,
\end{align}
where the Jacobian $J_T(\sampleX)$ is a $D\times D$ matrix of the partial derivatives of the transformation $T$:
\begin{align}
    J_T(\sampleX) = \left[\begin{array}{ccc}
         \frac{\partial T_1}{\partial x_1} & \cdots & \frac{\partial T_1}{\partial x_D}  \\
         \vdots & \ddots & \vdots \\
         \frac{\partial T_D}{\partial x_1} & \cdots & \frac{\partial T_D}{\partial x_D}
    \end{array}\right].
\end{align}

The transformation $T$ in a normalizing flow is defined partially through a neural network with parameters $\phi$, as will be described below, defining a density
\begin{align}\label{Eq:changeVariables}
    \probqphi{\sampleX} = \probU{T_\phi(\sampleX)} |\det J_{T_\phi}(\sampleX)|.
\end{align}
The goal is to find the parameters $\phi$ to maximize Eq.~\eqref{Eq:objective} if $\sampleX\sim\probp{\sampleX}$. The subindex  of $T_\phi$ will be omitted in the following, simply denoting the transformation of the neural network by $T$.

If the transformation is flexible enough, the flow could be used to evaluate any continuous density in $\mathbb{R}
^D$. In practice, however, the property that the composition of diffeomorphisms is a diffeomorphism is used, allowing to construct a complex transformation via composition of simpler transformations. Consider the transformation $T$ as a composition of simpler $T_k$ transformations:
\begin{align}
    T=T_K\circ \cdots \circ T_1.
\end{align}
Assuming $\sampleZ_0=\sampleX$ and $\sampleZ_K = \sampleU$, the forward evaluation and Jacobian are
\begin{align}
    \sampleZ_k &= T_k(\sampleZ_{k-1}),\; k=1:K,\\
    |J_T(\sampleX)| &= \left|\prod_{k=1}^K J_{T_k}(\sampleZ_{k-1})  \right|.
\end{align}
These two computations (plus their inverse) are the building blocks of a normalizing flow \cite{Rezende2015VariationalIW}. Hence, to make a transformation efficient, both operations have to be efficient. From now on forth, we will focus on a simple transformation $\sampleU=T(\sampleX)$, since constructing a flow from it is simply applying compositions.

To define a transformation satisfying both operations to be efficient, the transformation is broken down into autoregressive one-dimensional ones for each dimension of $\mathbb{R}^D$:
\begin{align}
u_i = \tau(x_i;\mathbf{h}_i)\text{ with } \mathbf{h}_i = c_i(\sampleX_{<i};\phi),
\end{align}
where $u_i$ is the $i$-th component of $\sampleU$ and $x_i$ the $i$-th of $\sampleX$.  $\tau$ is the \textit{transformer}, which is a one-dimensional diffeomorphism with respect to $x_i$ with parameters $\mathbf{h}_i$. $c_i$ is the $i$-th \textit{conditioner}, a neural network, which takes as input $\sampleX_{<i}=(x_1,x_2,\dots,x_{i-1})$, i.e., the previous components of $\sampleX$, and $\phi$ are the parameters of the neural network. The conditioner provides the parameters $h_i$ of the $i$-th transformer of $x_i$ depending on the previous components $\sampleX_{<i}$, defining implicitly a conditional density over $x_i$ with respect to $\sampleX_{<i}$. The transformer is chosen to be a differentiable monotonic function, since then it satisfies the requirements to be a diffeomorphism. The transformer also satisfies that it makes the transformation easily computational in parallel and decomposing the transformation in one dimensional autoregressive transformers allows the computation of the Jacobian to be trivial, because of its triangular shape. To compute the parameter $\mathbf{h}_i$ of each transformer, one would need to process a neural network with input $\sampleX_{<i}$ for each component, a total of $D$ times. 

\textit{Masked autoregressive neural networks} \cite{Germain2015MADEMA} enable to compute all the conditional functions simultaneously in a single forward iteration of the neural network. This is done by masking out, with a binary matrix, the connections of the $\mathbf{h}_i$-th output with respect to all the components with index bigger or equal to $i$, $\geq i$, making it a function of the $<i$ components.

The transformer can be defined by any monotonic function, such as affine transformations \cite{Papamakarios2017MaskedAF}, monotonic neural networks \cite{Huang2018NeuralAF,Cao2019BlockNA,Wehenkel2019UnconstrainedMN}, sum-of-squares polynomials \cite{Jaini2019SumofSquaresPF} or monotonic splines \cite{Mller2018NeuralIS,Durkan2019CubicSplineF,Durkan2019NeuralSF}. In this work we will focus on a specific implementation of monotonic splines, the Neural Spline Flows.

In their work on Neural Spline Flows \cite{Durkan2019NeuralSF}, Durkan et al. advocate for utilizing monotonic rational-quadratic splines as  transformers $\tau$, which are easily differentiable, more flexible than previous attempts using polynomials for these transformers, since their Taylor-series expansion is infinite, and are analytically invertible. 

The monotonic rational-quadratic transformers is defined by a quotient of two quadratic polynomial. In particular, the splines map the interval $[-B,B]$ to $[-B,B]$, and outside of it the identity function is considered. The splines are parametrized following Gregory and Delbourgo \cite{Gregory1982PiecewiseRQ}, where $K$ different rational-quadratic functions are used, with boundaries set by the pair of coordinates $\{(x^{(k)},u^{(k)}\}_{k=0}^K$, known as knots of the spline and are the points where it passes through. Note that $(x^{(0)},u^{(0)}) = (-B,-B)$ and $(x^{(K)},u^{(K)}) = (B,B)$. Additionally, we need $K-1$ intermediate positive derivative values, since the boundary points derivatives are set to 1 to match the identity function. 

Having this in mind, the conditioner given by the neural network outputs a vector $\mathbf{h}=[\mathbf{h}^w,\mathbf{h}^h,\mathbf{h}^d]$ of dimension $3K-1$ for the transformer $\tau$, $c_i(\sampleX_{<i};\phi)=\mathbf{h}_i$. $\mathbf{h}^w$ and $\mathbf{h}^h$ give the width and height of the $K$ bins, while $\mathbf{h}^d$ is the positive derivative at the intermediate knots. 

Stacking up many of these transformations, a highly flexible neural density estimator can be build and will be the one utilized during this work.

\subsection{Neural Spline Flows applied to the T2K Oscillation problem}
\label{Sec:NSFatT2K}

In this subsection we will explain how the likelihood for Bayesian inference is constructed. In the context of machine learning, likelihood-free inference, as stated in the abstract, refers to the task of performing such analysis when the densities are data-driven, but no explicit likelihood function can be constructed. This is the case for the near detector, where we have data for the different magnitudes but no analytical density available. Therefore, a density estimation for the near detector is performed through NSF. This near detector density is then combined with the analytical formula for neutrino oscillation for the far detector in order to obtain the likelihood for the experiment. By doing so, we are combining the potential of NSF with expertise of the particular problem.

We start by estimating the explicit density of the expected energy spectrum $E_\nu$ of the neutrinos, together with the momentum $p_\mu$ and angle $\theta_\mu$ of the measured muon without oscillations, obtaining $p(p_\mu,\theta_\mu,E_\nu)$, as measured by the near detector. This is done by learning the density using a NSF from the Monte Carlo data, generated as presented in Sec.~\ref{Sec:PhysicalSimulator}. The base density $\probU{\sampleU}$ used for Eq.~\eqref{Eq:changeVariables} is a three-dimensional standard normal distribution.

Having estimated the joint probability $p(p_\mu,\theta_\mu,E_\nu)$ of the initial distribution at the near detector, we need to construct the conditional density $p\left(p_\mu, \theta_\mu|\theta_\text{mix},\Delta m^2\right)$ of the observed magnitudes  given the oscillation parameters at the far detector in order to perform Bayesian inference. For this, we simply integrate the probability of not oscillating, $1-p_\text{osc}$, using Eq.~\eqref{Eq:POscillation}, over the energy spectrum of the joint distribution:
\begin{align}
&p\left(p_\mu, \theta_\mu|\theta_\text{mix},\Delta m^2\right) =  C\left(\theta_\text{mix},\Delta m^2\right) \cdot \\ 
&\phantom{p(p_\mu, \theta_\mu)}\int p(p_\mu,\theta_\mu,E_\nu) \big(1-p_\text{osc}(E_\nu,\theta_\text{mix},\Delta m^2)\big) dE_\nu,
\end{align}
where $C\left(\theta_\text{mix},\Delta m^2\right)$ is a constant of normalization computed after performing the integral. With this we have the probability of observing a single muon with momentum $p_\mu$ and angle $\theta_\mu$ after oscillating given the parameters $\theta_\text{mix}$ and $\Delta m^2$.

In order to take into account the number of observed samples, the extended likelihood \cite{PhysRevD.98.030001,Barlow1990} is used, modifying the likelihood with a Poisson count term to consider the expected number of events for a given set of parameters and the actual observed number:
\begin{align}\label{Eq:lklh}
L(\boldsymbol{\theta})=\frac{\left[\mu\left(\boldsymbol{\theta}\right)\right]^{n}}{n !} e^{-\mu\left(\boldsymbol{\theta}\right)} \prod_{i=1}^{n} p\left(\sampleX_{i} | \boldsymbol{\theta}\right).
\end{align}
In our case, the Poisson parameter $\mu\left(\boldsymbol{\theta}\right)$ is obtained by integrating the possible oscillations over all the energy spectrum, scaled to the initial number of particles $N_\text{ini}$:
\begin{gather}
\mu\left(\theta_\text{mix},\Delta m^2\right) = N_\text{ini} \times\\ 
\int\!\!\!\int\!\!\!\int\!\!\!p(p_\mu,\theta_\mu,E_\nu)
\Big(\!1-p_\text{osc}\left(E_\nu,\theta_\text{mix},\Delta m^2\right)\!\Big)dE_\nu dp_\mu d\theta_\mu.
\end{gather}
Hence, the extended likelihood we apply for the analysis is
\begin{align}
L\left(\theta_\text{mix},\Delta m^2\right) =& \frac{\left[\mu \left(\theta_\text{mix},\Delta m^2\right)\right]^{n}}{n !} e^{-\mu\left(\theta_\text{mix},\Delta m^2\right)}\times\\
&\prod_{i=1}^{n} p\left(p_\mu^{(i)}, \theta_\mu^{(i)}|\theta_\text{mix},\Delta m^2\right),
\label{Eq:Extlklh}
\end{align}
and the posterior for the parameters takes the form of
\begin{align}
   &p\left(\theta_\text{mix},\Delta m^2|\left\{p_\mu^{(i)}, \theta_\mu^{(i)}\right\}_{i=1}^n\right) \propto \\ &\hspace{12ex}L\left(\theta_\text{mix},\Delta m^2\right) p\left(\theta_\text{mix},\Delta m^2\right),\label{Eq:posterior}
\end{align}
with $p\left(\theta_\text{mix},\Delta m^2\right)$ the prior information before observing the events and $\left\{p_\mu^{(i)}, \theta_\mu^{(i)}\right\}_{i=1}^n$ the set of observed events.

\subsection{Reference analysis using an approximate un-binned likelihood} 
\label{Sec:unbinnedlklh}

The results of the experiments are validated using an approximate un-binned likelihood. The likelihood is computed following Eq.~\eqref{Eq:Extlklh}. To do so, event histograms  ($M(p_{\mu}^i,\theta_{\mu}^j|E^k)$) binned in muon momentum and angle given a neutrino energy are generated from the simulated data described in Sec.~\ref{Sec:PhysicalSimulator}. The oscillated probability is computed by reweighting the histogram content, using Eq.~\eqref{Eq:POscillation}, as 
 \begin{align}
 &M^\text{osc}(p_{\mu}^i,\theta_{\mu}^j|\theta_\text{mix},\Delta m^2) = \\ &\hspace{6ex}\sum_k M(p_{\mu}^i,\theta_{\mu}^j|E^k) p_\text{osc}(E^k,\theta_\text{mix},\Delta m^2),
\end{align}
and is interpolated linearly to reduce the effects due to the coarse binning:
\begin{align}
 M^\text{osc}(p_{\mu}^i,\theta_{\mu}^j|\theta_\text{mix},\Delta m^2) \rightarrow M^\text{osc}(p_{\mu},\theta_{\mu}|\theta_\text{mix},\Delta m^2).
\end{align}
 
The number of expected events ($\mu(\theta,\Delta m^2)$) is computed by summing the binned probability:
\begin{align}
 \mu(\theta_\text{mix},\Delta m^2) = \sum_{i,j} M^\text{osc}(p_{\mu}^i,\theta_{\mu}^j|\theta_\text{mix},\Delta m^2). 
\end{align}
The probability obtained by normalizing the oscillated maps takes the form
\begin{align}
p\left( p_{\mu},\theta_{\mu} | \theta,\Delta m^2\right) = \frac{ M^\text{osc}(p_{\mu},\theta_{\mu}|\theta_\text{mix},\Delta m^2)}{ \mu(\theta_\text{mix},\Delta m^2)  } .
\end{align}

The likelihood probability $L(\theta_\text{mix},\Delta m^2)$ is finally computed for discrete values of $\theta_\text{mix}$ and $\Delta m^2$. Those values are distributed in a grid identical to the one used for the NSF approach for proper comparison between both methods. We have tested the results with different number of initial bins in energy, momentum and angle to find a good compromise between stability of the result and speed. We call this cross-check method in what follows the \textit{Hist} method and it will be used as a reference for the NSF calculations.

\section{Experiments}
\label{Sec:Experiments}

The methodology is tested on experiments of simulated neutrino oscillations according to the T2K experiment. For this, ten different set of observed events are constructed (see Appendix~\ref{Sec:experimentSimulation}), with additional Monte Carlo (MC) data used to fit both the NSF and construct the un-binned likelihood. The training of the NSF to fit the density of $(p_\mu,\theta_\mu,E_\nu)$ is shown and verified in Sec.~\ref{Sec:NSFvalidation}. Afterwards, in Sec.~\ref{Sec:Results}, the inference on the ten experiments are performed utilizing both for NSF and the un-binned likelihood, discussing the findings of both methodologies and the possible bias introduced by them.

All events were generated as defined in Sec.~\ref{Sec:PhysicalSimulator}, using the neutrino flux energy spectrum from the T2K collaboration \cite{PhysRevD.87.012001} to produce the energy of the incoming neutrino $E_\nu$, and the NEUT event generator \cite{Hayato:2009zz} to compute the momentum $p_\mu$ and angle $\theta_\mu$ of the resulting muon, forming a triplet of  $(p_\mu,\theta_\mu,E_\nu)$, describing an implicit density of these three magnitudes.

\subsection{Training and validation of the NSF}
\label{Sec:NSFvalidation}

In order to apply the methodology described in Sec.~\ref{Sec:NSFatT2K}, we start by estimating the density $p(p_\mu,\theta_\mu,E_\nu)$ using a Neural Spline Flow on $\approx15$M MC events as the training set to fit the parameters of the flow. As is standard procedure in machine learning, an additional set of $\approx4$M events are used for validation of the model outside of the training set, i.e., to check the integrity of the models for data not seen to fit the parameters.

\begin{figure}[!thp]
\begin{center}
\includegraphics[width=\columnwidth]{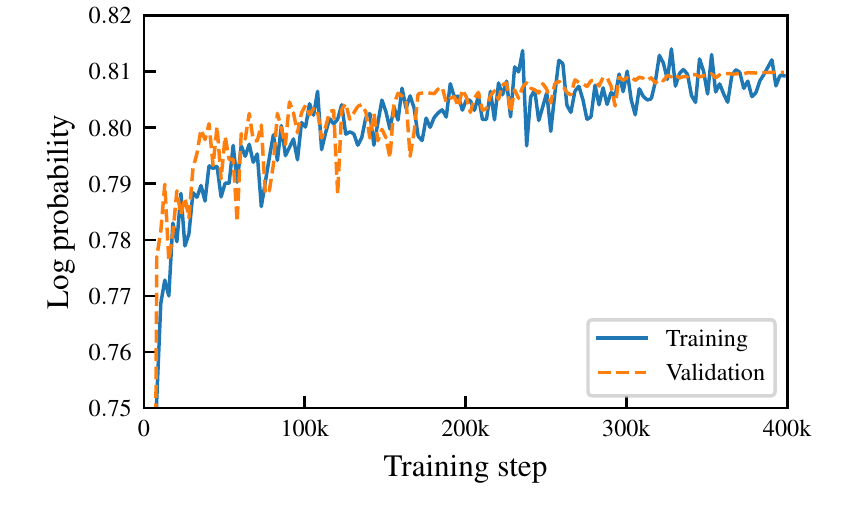}
\caption{Neural Spline Flow training log probability for estimating $p(p_\mu,\theta_\mu,E_\nu)$ for training (solid) and validation (dashed) sets, as shown by Eq.~\eqref{Eq:objective}. For the validation, the log probability stabilizes during the training to converge to a certain value which depends on the architecture of the network.}
\label{Fig:trainLog}
\end{center}
\end{figure}

Because of the similarities regarding dimension of the data and \# of samples, but with a simpler structure, we have chosen the hyperparameters almost as the \textsc{Power} data-set in Appendix~B from \cite{Durkan2019NeuralSF}, i.e., \textsc{Adam} optimizer \cite{Kingma2014AdamAM} with learning rate 0.0005, batch size 512, training steps 400k, flow steps 5, transform blocks 5, hidden features 128 and bins 8. Additionally, the learning rate was decreased during the training using a cosine scheduler to ensure stabilization at the end of the training procedure. As shown in Fig.~\ref{Fig:trainLog}, the validation set stabilizes at the end of the training and appears to converge for the selected architecture of the network.

\begin{figure}[htp]
\begin{center}
\includegraphics[width=\columnwidth]{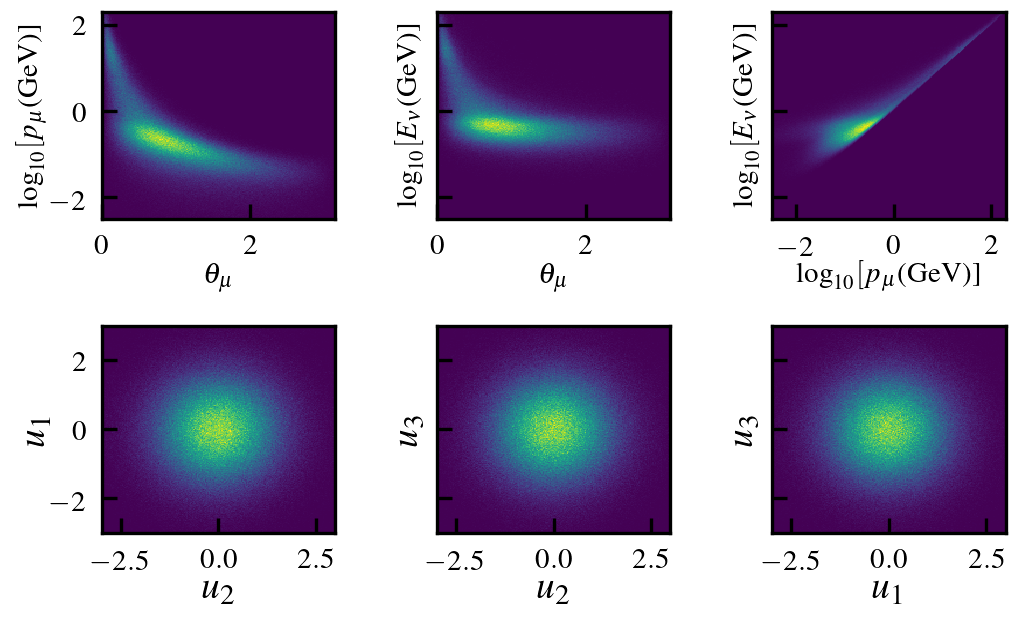}
\caption{Two dimensional histograms of samples from the initial distribution of $(p_\mu,\theta_\mu,E_\nu)$ (top) and the transformed data $\mathbf{u}=(u_1,u_2,u_3)$ (bottom) under $T$ according to Eq.~\eqref{Eq:changeVariables}. If the transformation $T$ is approximated properly, $\mathbf{u}$ should follow a three-dimensional standard normal distribution, as is depicted qualitatively in this Figure.}
\label{Fig:xtoz}
\end{center}
\end{figure}

To ensure that the transformation $T$ of Eq.~\eqref{Eq:changeVariables} was found properly by the NSF, consider the transformed data $\sampleU$ for a new set of $\approx 1$M MC events, never seen before by the algorithm. If $T$ is correctly approximated, $\sampleU$ should follow a three-dimensional standard normal distribution, as is shown qualitatively in Fig.~\ref{Fig:xtoz}.

For this, a $\chi^2$-test was performed over a binning of $50\times50\times50$ in the domain $[-3,3]^3$ of the transformed data $\sampleU$ to test the goodness-of-fit to a three-dimensional standard normal distribution, obtaining a $p$-value of 0.3062. With this, we can assume that the transformed data $\sampleU$ corresponds to samples from such base distribution, justifying that the transformation $T$ was properly found by the NSF, hence allowing to evaluate $p(p_\mu,\theta_\mu,E_\nu)$ accurately through it.

\subsection{Inference results}
\label{Sec:Results}

To test the performance of obtaining the posterior according to Sec.~\ref{Sec:NSFatT2K}, ten different observation sets were constructed, as explained in Appendix~\ref{Sec:experimentSimulation}, with five different mixing angles $\theta_\text{mix}$ and difference in mass squared $\Delta m^2$. 

For each of the five combinations of parameters, low and high statistics (number of observed events) experiments were performed. Low statistics are of the order of the real observed samples at T2K and used to assure its performance when a small number of events is dealt with. High statistics (two orders of magnitude larger number of observed events compared to the usual expected number in the low statistics case) allow us to check the agreement between traditional binning methodology with a large number of very fine bins and the un-binned NSF posterior.  

\begin{table}[ht!]
    \caption{Posterior inference of ten different experiments, alternating between low and high number of observed events. For the inferred parameters using NSF and the un-binned histogram approximation, \textit{Hist}, the 95\% confidence level was computed using the 1-dimensional marginalized densities of each parameter. $\theta_\text{mix}$ is given in rad and $\Delta m^2$ in $\times 10^{-3}\mathrm{eV}^2$.}
    \label{Tab:Posteriors}
\centering
\small{
\begin{tabular}{c|c|c|c|c|c}
Exp.& $N_\text{obs}$ & Parameter & True  & NSF & \textit{Hist} \\
\#  &                &        & values &  95 \% C.L.  &   95 \% C.L. \\  
\hline
\rule[-6pt]{0pt}{18pt} 1 & 506 & $\theta_\text{mix}$ & 0.7594 & $0.785^{+0.055}_{-0.056}$ & $0.785^{+0.055}_{-0.056}$\\
\rule[-6pt]{0pt}{0pt}    &    & $\Delta m^2$ & 2.463 & $2.440^{+0.091}_{-0.085}$ & $2.446^{+0.092}_{-0.087}$ \\
\hline
\rule[-6pt]{0pt}{18pt} 2 & 49672 & $\theta_\text{mix}$ & 0.7594 & $0.751^{+0.074}_{-0.006}$ & $0.754^{+0.069}_{-0.007}$\\
\rule[-6pt]{0pt}{0pt}    &    & $\Delta m^2$ & 2.463 & $2.464^{+0.008}_{-0.012}$ & $2.467^{+0.007}_{-0.012}$ \\
\hline
\rule[-6pt]{0pt}{18pt} 3 & 532 & $\theta_\text{mix}$ & 0.7353 & $0.71^{+0.18}_{-0.03}$ & $0.71^{+0.18}_{-0.03}$\\
\rule[-6pt]{0pt}{0pt}    &    & $\Delta m^2$ & 2.463 & $2.46^{+0.10}_{-0.11}$ & $2.46^{+0.10}_{-0.11}$ \\
\hline
\rule[-6pt]{0pt}{18pt} 4 & 49646 & $\theta_\text{mix}$ & 0.7353 & $0.738^{+0.099}_{-0.006}$ & $0.739^{+0.097}_{-0.006}$\\
\rule[-6pt]{0pt}{0pt}    &    & $\Delta m^2$ & 2.463 & $2.458^{+0.009}_{-0.011}$ & $2.461^{+0.009}_{-0.011}$ \\
\hline
\rule[-6pt]{0pt}{18pt} 5 & 493 & $\theta_\text{mix}$ & 0.6847 & $0.65^{+0.28}_{-0.02}$ & $0.65^{+0.28}_{-0.02}$\\
\rule[-6pt]{0pt}{0pt}    &    & $\Delta m^2$ & 2.463 & $2.554^{+0.120}_{-0.127}$ & $2.563^{+0.122}_{-0.128}$ \\
\hline
\rule[-6pt]{0pt}{18pt} 6 & 49665 & $\theta_\text{mix}$ & 0.6847 & $0.682^{+0.210}_{-0.003}$ & $0.683^{+0.208}_{-0.004}$\\
\rule[-6pt]{0pt}{0pt}    &    & $\Delta m^2$ & 2.463 & $2.479^{+0.008}_{-0.014}$ & $2.482^{+0.009}_{-0.014}$ \\
\hline
\rule[-6pt]{0pt}{18pt} 7 & 506 & $\theta_\text{mix}$ & 0.7353 & $0.73^{+0.14}_{-0.04}$ & $0.73^{+0.14}_{-0.04}$\\
\rule[-6pt]{0pt}{0pt}    &    & $\Delta m^2$ & 2.363 & $2.347^{+0.097}_{-0.095}$ & $2.353^{+0.096}_{-0.099}$ \\
\hline
\rule[-6pt]{0pt}{18pt} 8 & 50145 & $\theta_\text{mix}$ & 0.7353 & $0.734^{+0.108}_{-0.006}$ & $0.735^{+0.106}_{-0.006}$\\
\rule[-6pt]{0pt}{0pt}    &    & $\Delta m^2$ & 2.363 & $2.365^{+0.010}_{-0.011}$ & $2.368^{+0.009}_{-0.012}$ \\
\hline
\rule[-6pt]{0pt}{18pt} 9 & 481 & $\theta_\text{mix}$ & 0.7353 & $0.785^{+0.060}_{-0.061}$ & $0.785^{+0.060}_{-0.061}$\\
\rule[-6pt]{0pt}{0pt}    &    & $\Delta m^2$ & 2.663 & $2.620^{+0.086}_{-0.087}$ & $2.626^{+0.087}_{-0.089}$ \\
\hline
\rule[-6pt]{0pt}{18pt} 10 & 49710 & $\theta_\text{mix}$ & 0.7353 & $0.741^{+0.094}_{-0.005}$ & $0.743^{+0.089}_{-0.005}$\\
\rule[-6pt]{0pt}{0pt}    &    & $\Delta m^2$ & 2.663 & $2.659^{+0.007}_{-0.011}$ & $2.662^{+0.008}_{-0.010}$ 
\end{tabular}
}
\end{table}

Table.~\ref{Tab:Posteriors} shows the ten experiments, with the number of observed samples, the true parameters and the results using the NSF (Sec.~\ref{Sec:NSFatT2K}). Additionally, a result using an approximate un-binned likelihood, denoted by \textit{Hist} (Sec.~\ref{Sec:unbinnedlklh}), is also displayed, which would correspond to the limit case when histograms can be performed with a large number of bins to behave like an un-binned estimation. Since Bayesian inference is used, the inference on the parameters describes a density function according to Eq.~\eqref{Eq:posterior}. The central value shown for each parameter is the one that maximizes the joint posterior density. The uncertainty is then computed by marginalizing in the 2-dimensional density one of the parameters to obtain the 1-dimensional one of the other, and finding the interval such that for a $1-\alpha$ confidence level (CL), $\alpha/2$ of the density is found on each side. This is done for both NSF posterior and \textit{Hist} posterior.

In general, the results of both methodologies agree, with slight fluctuations in the confidence levels. The difference could come from the NSF not learning perfectly the density of the points, from the interpolation done by the \textit{Hist} method introducing wrong approximations or from intrinsic biases, which will be discussed at the end o this subsection.

\begin{figure*}[tp]
\begin{center}
\includegraphics[width=\columnwidth]{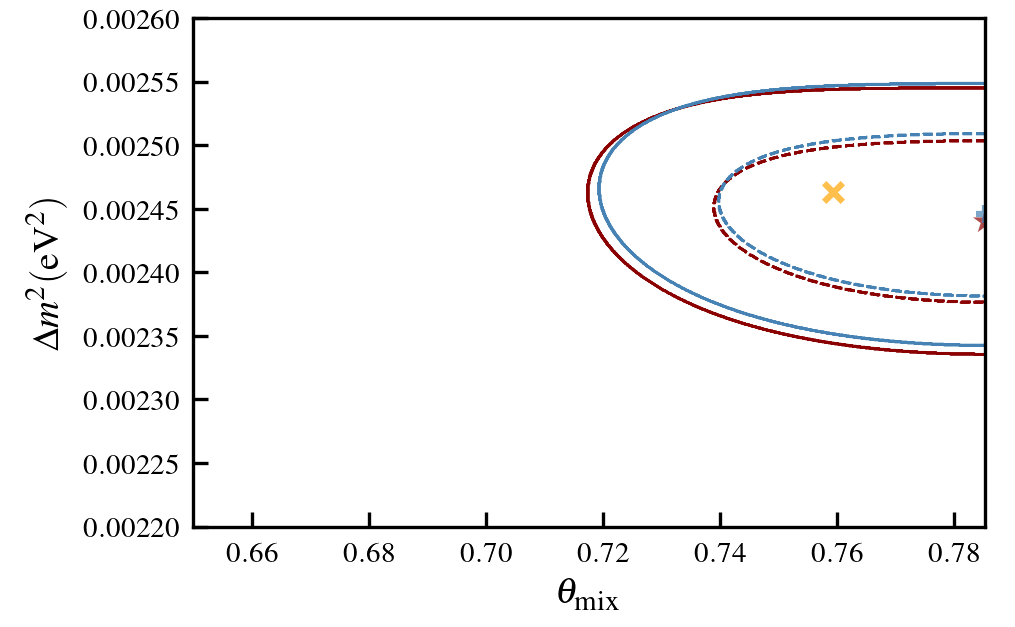}
\includegraphics[width=\columnwidth]{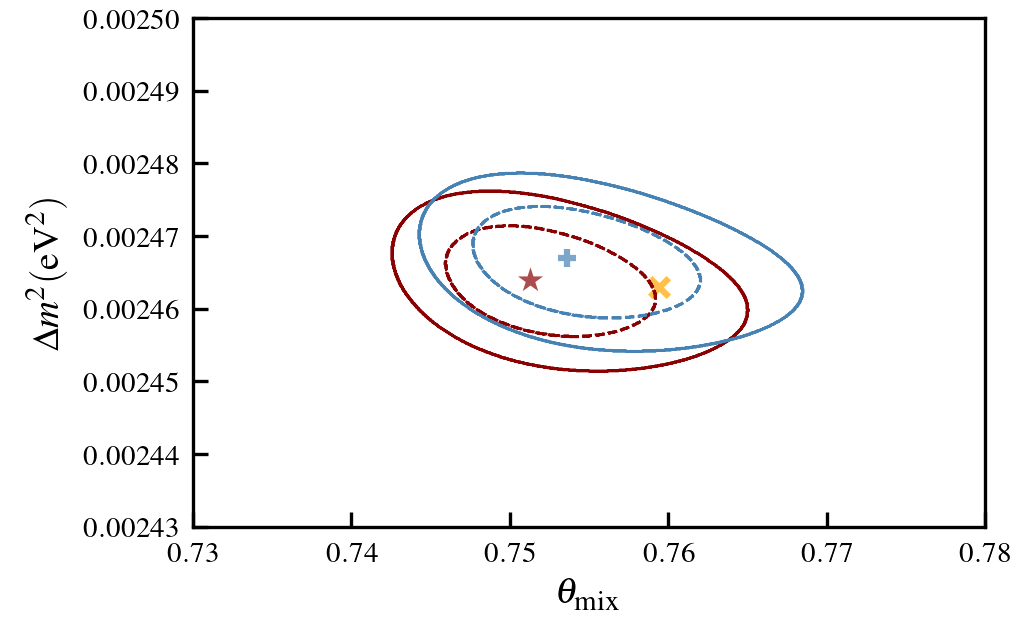}
\includegraphics[width=\columnwidth]{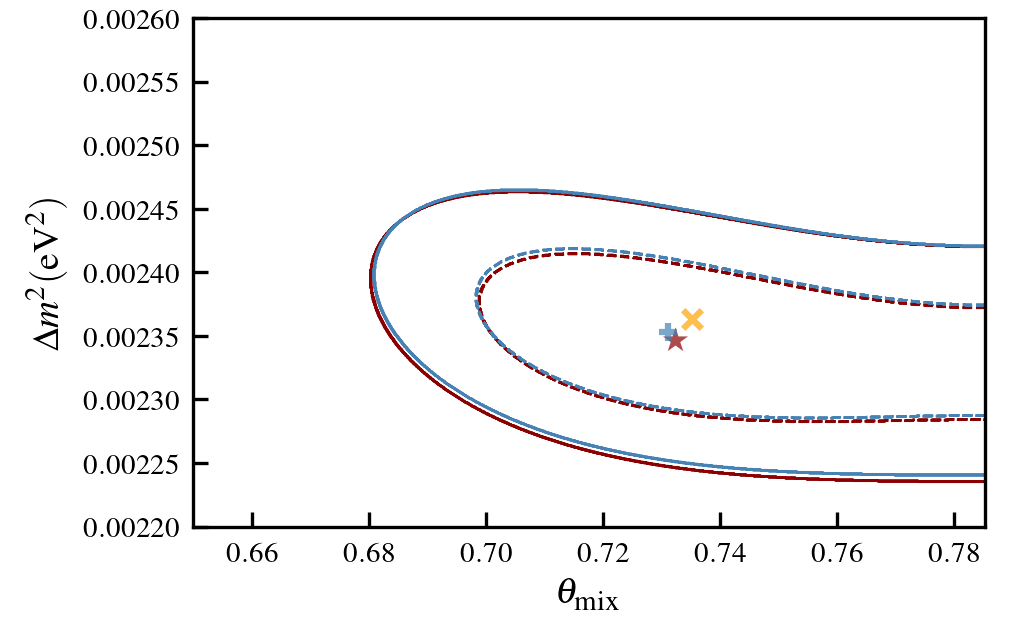}
\includegraphics[width=\columnwidth]{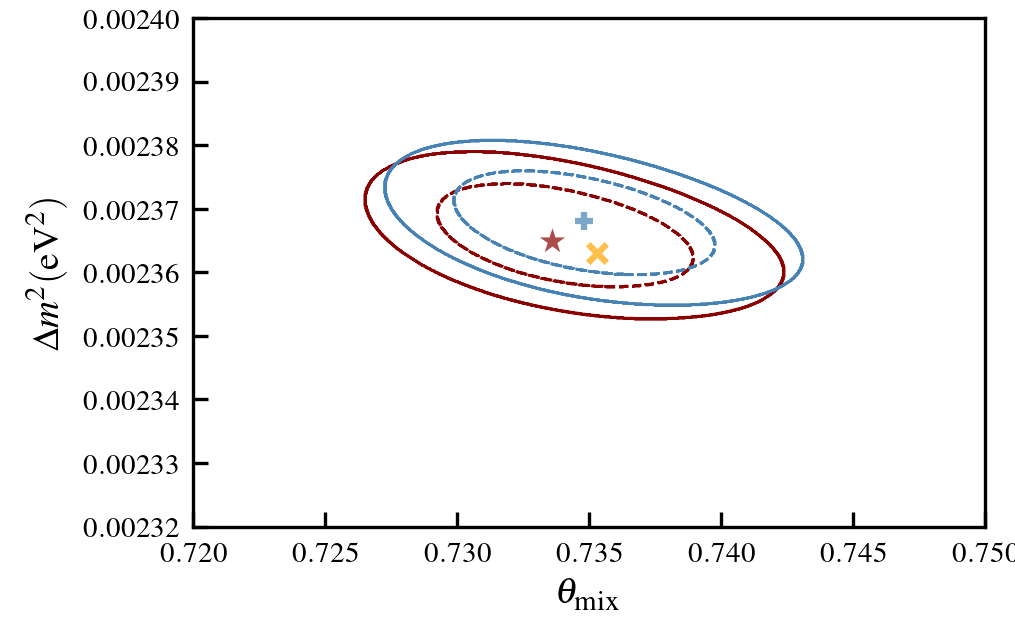}
\caption{Highest Posterior Density (HPD) curves for both NSF and \textit{Hist} posteriors of experiments 1 (top left), 2 (top right), 7 (bottom left) and 8 (bottom right), together with best fit of each posterior. Red (Blue) lines indicate 68 \% (dashed) and 95\% (continuous) HPD curves for the NSF (\textit{Hist}) method. Orange x-crosses indicate the true parameter used to generate the observed events. Red stars (blue +-crosses) indicate the best fit for the NSF (\textit{Hist}) method.  A clear overlap can be found in experiments of low statistics (left) and a slight fluctuation on large statistics (right). Notice a change of scale in the high statistics plots.}
\label{Fig:posteriors}
\end{center}
\end{figure*}

Additionally, in order to visualize the agreement in 2-dimensions in Fig.~\ref{Fig:posteriors}, the Highest Posterior Density (HPD) curves \cite{1390921} of 68\% and 95\%, together with the best fit (highest posterior value) were computed for experiments 1 (top left), 2 (top right), 7 (bottom left) and 8 (bottom right). In both HPD regions a clear overlap for low statistics, and slight fluctuations on larger statistics can be observed. When comparing the difference in area size, the relative difference of the \textit{Hist} method with respect to NSF through the ten experiments is $3.1\pm1.9\%$, showing that the areas agree within 2-$\sigma$ on average.

In Fig.~\ref{Fig:posteriors}, one observes that the areas, even being similar in size, are slightly shifted one from another. Empirical experiments of computing the posterior using actual discrete binning show that, by using a denser binning, its posterior was shifting towards the NSF result. To measure the bias ($\hat{\theta}-\theta$) of the estimated parameters $(\hat{\theta}_\text{mix},\hat{\Delta m}^2)$ by both methods, an observation of  $N_\text{obs}\approx500$k events were used for the two sets of parameters used in Fig.~\ref{Fig:posteriors}. Results are summarized in Tab.~\ref{Tab:bias} where $(\hat{\theta}_\text{mix},\hat{\Delta m}^2)$ are taken as the maximum value of the posterior probability obtained in each set of parameters. Tab.~\ref{Tab:bias} shows that NSF has a significant smaller bias compared to the \textit{Hist} method. This explains the discrepancy in the plots, aside from justifying a better performance by the NSF method. The \textit{Hist} bias comes from binning and interpolation approximations while for NSF the bias may come from the density estimation for the near detector not being perfect. In real experiments this bias can be estimated using a representative MC data set, called the \textit{Asimov data set}.

\begin{table}[htbp]
    \centering
    \caption{Bias computation for the posterior densities of the parameters in Fig.~\ref{Fig:posteriors}. The estimators $(\hat{\theta}_\text{mix},\hat{\Delta m}^2)$ are taken as the maximum value of the posterior obtained in each set of parameters for a high-statistical experiment ($N_\text{obs}\approx500$k), and the bias is defined as $\hat{\theta}-\theta$. NSF shows a significant reduction of bias compared to the \textit{Hist} method. $\theta_\text{mix}$ is given in rad and $\Delta m^2$ in $\times 10^{-3}\mathrm{eV}^2$.}
    \label{Tab:bias}
    \begin{tabular}{cccc}
    Parameter & True & NSF & \textit{Hist} \\
    & values & bias & bias\\
    \hline
    $\theta_\text{mix}$ & 0.7594 & 0.0020 & 0.0087\\
    $\Delta m^2$ & 2.463 & -0.0012 & 0.0031\\
    \hline
    $\theta_\text{mix}$ & 0.7353 & -0.00050 & 0.00410\\
    $\Delta m^2$ & 2.363 & -0.00063 & 0.00675
    \end{tabular}
\end{table}

Both quantitative, Table~\ref{Tab:Posteriors}, and qualitative, Fig.~\ref{Fig:posteriors}, show that NSF indeed provide a tool to perform likelihood-free inference on physical simulators such as the one of the T2K experiment, in agreement with un-binned likelihood approaches as we have compared it to, but with less bias as shown in Tab.~\ref{Tab:bias}.

\section{ Modelling systematic uncertainties } 
\label{Sec:RealCase}

 A comprehensive description of the problem is beyond the scope of this letter, but we will sketch possible implementation alternatives. In experiments, near detector neutrino interactions data is used to constrain uncertainties in cross-section models ($\hat\phi_\text{xsect}$), neutrino flux ($\hat\phi_\text{flux}$) and detector smearing and efficiency ($\hat\phi_\text{det}$). Comparing the experimental data from the near detector to the Monte Carlo model, experiments obtain the distribution for the parameters, $p(\hat \phi|\text{ND})$.
 The uncertainty parameters ($\hat\phi_{i}$) are applied to the far detector to predict the data distributions in absence of oscillations. The transport of constrained uncertainties are done either by traditional covariance matrices or by a more sophisticated Markov Chain Monte Carlo that easily accounts for non Gaussian probability distributions. The uncertainties are then marginalize or profiled to propagate the uncertainties to the oscillation analysis. 
 
  The proposed method can be used in different ways in this analysis framework. The simplest approach is to obtain the $p(\hat \phi|\text{ND})$ that includes all possible parameter correlations. In this case, the model provides at the same time a simple way to generate Monte Carlo to sample the distributions. This method, using Gaussian {\it base densities}, will easily learn the nuisance parameters probability density function which is expected to be close to Gaussian. The next level of complexity is to learn, as we have done in this example, the probability density function but adding nuisance parameters, $p(p_\mu,\theta_\mu,E_\nu,\hat\phi)$. This implementation will allow to perform unbinned likelihoods as described in this letter. The more complex and inclusive approach is to avoid the intermediate nuisance parameter density function ($p(\hat \phi|\text{ND})$) description and model both near and far detectors with a set of common uncertainties. The advantage of this final description is that all the analysis is carried out in a single fit avoiding the description of hundreds of uncertainties ($\hat\phi$).

\section{Summary}

In this work we have presented the viability of a likelihood-free inference methodology through Neural Spline Flows on a simplified neutrino oscillation problem at the T2K experiment. 
A brief introduction to normalizing flows as density estimators from data samples is performed, making emphasis on the particular implementation of Neural Spline Flows. We developed a framework to use this estimation of the density from data taken at the near detector in order to perform inference of the oscillation parameters at the far detector for a simplified 2-flavour neutrino problem, allowing to perform exact inference if the density is properly estimated. This method provides potential advantages over traditional binned histogram methods, specially when the statistics is low as is the case in the T2K experiment. An un-binned alternative formulated by interpolating the histogram method was constructed to check the results.  Additionally the integrity of the learned density was thoroughly verified through different statistical and empirical tests. The results obtained using the Neural Spline Flow methodology and the un-binned likelihood methodology show results which are in agreement with each other in the estimation of the statistical errors. The alternative method is not refined enough and it shows larger bias in the estimated parameters. The results presented in this letter open new possibilities to use similar likelihood-free neural network inference for more complex statistical analyses.

\begin{acknowledgments}
The authors have received funding from the Spanish Ministerio de Econom\'{i}a y Competitividad (SEIDI-MINECO) under Grants No.~FPA2016-77347-C2-2-P and SEV-2016-0588, from the Swiss National Foundation Grant No. 200021\_85012 and acknowledge the support of the Secretariat of Universities and Research of the Department of Business and Knowledge of the Generalitat of Catalonia. The authors also are indebted to the Servei d'Estad\'istica Aplicada from the Universitat Aut\`onoma de Barcelona for their continuous support in the preparation of this document. 
\end{acknowledgments}

\bibliography{references}

\appendix
\section{Experiment simulation}
\label{Sec:experimentSimulation}

In this Appendix, we will describe how observed events $(p_\mu,\theta_\mu,E_\nu)$ are generated for the different experiments. Because of statistical fluctuations due to an event oscillation with certain probability (following Eq.~\eqref{Eq:POscillation}), the exact number of observed events cannot be determined beforehand, but can be approximately be chosen. In order to generate the observations of a fixed set of parameters $(\theta_\text{mix},\Delta m^2)$, the procedure is the following:
\begin{enumerate}
    \item Choose an approximated number of observed events $N_\text{approx}$. 
    \item Generate a MC event $(p_\mu,\theta_\mu,E_\nu)$, make the event oscillate according to its energy $E_\nu$ and accept it with probability given by Eq.~\eqref{Eq:POscillation}, until accepting a total of $N_\text{approx}$ events. This takes a stochastic number of $N_\text{init}$ tries, which corresponds to the number of initial events before oscillating.
    \item Generate new $N_\text{init}$ number of MC events, and accept them with probability according to Eq.~\eqref{Eq:POscillation}. The accepted samples form the observation set, with a number of $N_\text{obs}$ samples, which is similar to $N_\text{approx}$.
\end{enumerate}
To summarize, given a fixed set of parameters $(\theta_\text{mix},\Delta m^2)$, we determine an approximate initial number $N_\text{init}$ of events before oscillating needed to generate $N_\text{approx}$ oscillated observations. However, because this is a stochastic procedure and we stopped exactly when $N_\text{approx}$ were generated, the procedure has to be repeated with a fixed number of $N_\text{init}$ tries, giving us $N_\text{obs}\sim N_\text{approx}$ observed events.

Two kind of experiments were performed for five different set of parameters: one of low statistics (around 500 observations), with a number of the order of the real number of observed events in T2K, and one of high statistics (around 50k observations), to see how the algorithm behaves and compares to traditional methods when having a larger amount of data available.

The parameters of the initial experiments 1 and 2 in Sec.~\ref{Sec:Results} are chosen following the best fit value of $\sin^2 \theta_{23}$ and $\Delta m^2$ from \cite{PhysRevLett.121.171802}. The parameter $\theta_\text{mix}$ is computed as 
\begin{equation*}
    \theta_\text{mix} = \frac{\pi}{2}-\arcsin\sqrt{\sin^2 \theta_{23}},
\end{equation*}
where we have used the fact that Eq.~\ref{Eq:POscillation} is symmetric over $\theta_\text{mix} = \pi/4$ and fixed it in the interval $[0,\pi/4]$. For the best fit value, maximal mixing is almost achieved ($\sin 2\theta_\text{mix}=0.9986$). Experiments 3 and 4 (5 and 6) follow the same procedure, but choosing $\sin^2 \theta_{23}$ within 1 (2) $\sigma$ of the best fit. Experiments 7-10 the mixing angle has the value of experiments 3 and 4, but $\Delta m^2$ is changed in order to explore the behaviour on the parameter space.
\end{document}